\documentclass[conference]{IEEEtran}
\def\IEEEtitletopspace{18pt}
\IEEEoverridecommandlockouts
\usepackage{cite}
\usepackage{amsmath,amssymb,amsfonts}
\usepackage{algorithmic}
\usepackage{graphicx}
\usepackage{textcomp}
\usepackage{xcolor}
\usepackage{booktabs}
\usepackage{url}
\def\BibTeX{{\rm B\kern-.05em{\sc i\kern-.025em b}\kern-.08em
    T\kern-.1667em\lower.7ex\hbox{E}\kern-.125emX}}

\usepackage{float}
\usepackage{hyperref}
\usepackage{breakurl}
\usepackage{dblfloatfix}

\begin{document}

\title{EEG-Based Framework for Reflexive and Perceptual
Assessment in CLIS: Preliminary Study in Healthy
Volunteers

\author{Nicoli Leal$^{1}$, Rute Bettencourt$^{2}$, Urbano J. Nunes$^{3}$, Gabriel Pires$^{4}$ 
\thanks{*This work was supported by the Portuguese Foundation for Science and Technology (FCT) through the BCI4ALL R\&D Project (COMPETE2030-FEDER-00842800, 2023.17977.ICDT) and by the R\&D Unit UID/00048, Instituto de Sistemas e Robótica – Coimbra (ISR-UC). Rute Bettencourt has been supported by the FCT Ph.D. scholarship 2023.03995.BD (DOI: 10.54499/2023.03995.BD).}
\thanks{$^{1}$Nicoli Leal  is with the Institute of Systems and Robotics, University of Coimbra, Portugal
        {\tt\small nicoli.lealo@isr.uc.pt}}%
\thanks{$^{2}$Rute Bettencourt is with Institute of Systems and Robotics, University of Coimbra, Portugal
        {\tt\small rute.bettencourt@isr.uc.pt }}%
\thanks{$^{3}$Urbano J. Nunes is with Institute of Systems and Robotics, University of Coimbra, Portugal, and also with Department of Electrical and Computer Engineering, University of Coimbra, Portugal
        {\tt\small urbano@deec.uc.pt}}%
\thanks{$^{4}$Gabriel Pires is with the Institute of Systems and Robotics, University of Coimbra, Portugal, and also with the Engineering Department of the Polytechnic Institute of Tomar, Portugal
        {\tt\small gpires@isr.uc.pt}}%
}

}

\maketitle

\begin{abstract}
 Despite the general assumption that completely locked-in state (CLIS) patients remain conscious and aware of their environment, the effectiveness of brain-computer interfaces (BCIs) in facilitating communication has been limited, as reported both in the literature and in our own findings. This limitation is likely attributable to impairments in executive functions, working memory, and vigilance, which appear to hinder the establishment of reliable BCI-based communication. The main goal of this research is to develop a neurophysiological report designed to support the evaluation of the cognitive state of these individuals and determine their ability to interact with BCIs. To achieve this, we designed a set of paradigms to assess CLIS patients at the reflexive and perceptual levels, based on neural responses associated with sensory and perceptual processing, including Mismatch Negativity (MMN), Steady State Auditory Evoked Potential (SSAEP), and Steady State Visual Evoked Potential (SSVEP). Pilot testing with five healthy participants demonstrates the feasibility of generating a neurophysiological report for cognitive assessment at both levels. 
\end{abstract}


\begin{figure*}[t]
\centering \includegraphics[height=0.17\linewidth]{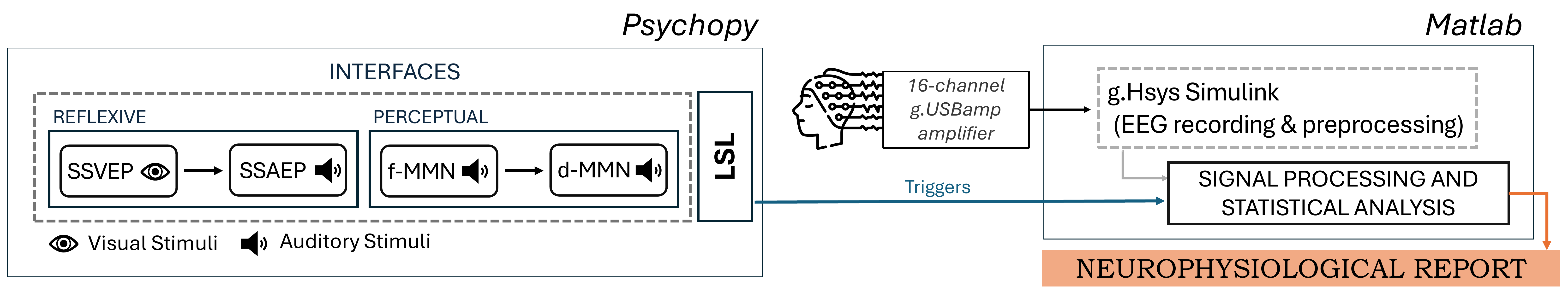}
\caption{
Overview of the experimental framework. On the left, PsychoPy implements reflexive and perceptual paradigms using Steady-State Visual and Auditory Evoked Potentials (SSVEP and SSAEP), as well as Mismatch Negativity paradigms in both the frequency (f-MMN) and duration (d-MMN) domains, with event triggers sent via LSL. On the right, EEG is recorded and processed in Matlab, yielding as a final contribution the neurophysiological report.
} 

\vspace{-0.5cm}
\label{fig:pipeline} 
\end{figure*}

\section{Introduction}
The Completely Locked-In State (CLIS) is characterized by total motor paralysis, including the loss of eye movements and facial expressions. Patients in this state are unable to express themselves and interact with their environment, while it is believed that they are aware of their surroundings  \cite{b1}. 

Brain-computer interfaces (BCIs) can establish a communication channel based solely on brain activity, bypassing conventional motor pathways. However, enabling communication in CLIS patients remains highly challenging, with the best outcomes limited to basic binary (yes/no) responses \cite{b1,b2}, typically using P300 or motor imagery paradigms. In a previous study \cite{b3} spanning ten months with a CLIS patient, we observed high EEG variability, possibly resulting from fluctuations in cognitive processes, attention and alertness, which contributed to unpredictable BCI performance. These findings reinforce a central concern highlighted in \cite{b4}, that the actual state of consciousness in CLIS patients remains uncertain. Consequently, there is no assurance they can reliably use a BCI. Detecting functional levels is essential for accurate diagnosis, treatment decisions, and assessing whether future communication pathways are viable \cite{b5}.

Awareness and functional capacity in CLIS patients have been investigated in only a few studies. Most of these works have focused on characterizing awareness using power spectral features during resting-state EEG \cite{b44}. In \cite{b6}, arousal levels were estimated via biomarkers such as the power law exponent (PLE) and Lempel-Ziv complexity (LZC), also in resting-state, to predict BCI performance. In our own study \cite{b45}, we also found that PLE correlated with BCI performance, though it was assessed during online BCI operation. Additionally, we showed that EEGNet, a deep learning model, could help predict BCI performance from EEG data. In \cite{b10}, spectral features recorded during resting state and sleep, as well as auditory and somatosensory stimulation, were used to provide a neurophysiological characterization. The P300 ERP was used in \cite{b7} to detect voluntary responses as a sign of consciousness. However, the absence of a P300 does not confirm lack of awareness, as patients may respond to passive but not active stimuli due to impaired comprehension or task-following ability \cite{b8}. Accurately assessing consciousness in CLIS patients remains critically important.

This study proposes a systematic approach to assess awareness in CLIS patients and assess their potential to benefit from BCI-based communication. We propose a two-level hierarchical framework targeting sensory processing (reflexive level) and stimulus discrimination (perceptual level), both essential for BCI usability. To this end, we designed EEG paradigms to elicit neural responses at each level: two mismatch negativity (MMN) variants (frequency and duration deviance), and steady-state visual (SSVEP) and auditory (SSAEP) evoked potentials. The goal is to define normative neurophysiological reference ranges, similar to statistical norms used in clinical blood tests, and to associate these ranges with predictive indicators of BCI performance. This structured EEG-based framework is currently being tested in a pre-control group of healthy volunteers and will support a follow-up study involving a larger control group matched by age and varied educational backgrounds. The framework may also aid in detecting cognitive motor dissociation (CMD) in disorders of consciousness (DoC).

\section{Methods}

\subsection{Software Environment and Stimulus Presentation}

All experimental paradigms were implemented using PsychoPy, with real-time event markers sent to MATLAB via the Lab Streaming Layer (LSL). High-speed Simulink (g.HIsys) was used for real-time data acquisition and processing. Auditory stimuli were presented binaurally via earphones, while visual stimuli were displayed on a computer monitor with a resolution of 1440 x 900 pixels and a refresh rate of 60 Hz (ASUS VW193D).

\subsection{EEG Acquisition and Participants}

EEG signals were acquired using a 16-channel g.USBamp amplifier (g.tec, Austria). The acquisition setup included a g.GAMMAcap2 EEG cap with active electrodes placed according to the extended international 10-20 system at the following positions: FPz, Fz, FCz, T7, C3, Cz, C4, T8, CPz, P3, Pz, P4, POz, O1, Oz, and O2. The reference electrode was placed on the right earlobe, and the ground electrode was located at the AFz site. 

The EEG data were recorded at a sampling rate of 256 Hz and signals were filtered online with a bandpass filter between 0.1–100 Hz for the SSVEP and SSAEP paradigms and between 0.1–30 Hz for the MMN paradigms. Additionally, a 50 Hz notch filter was applied across all paradigms. 

The study included five healthy control participants (2 female, 25.2 $\pm$ 2.5 years), all with normal or corrected-to-normal vision. This study was approved by the Ethical Committee for Health of the University Hospital Center of S. João, Porto.

\subsection{Neural Paradigms: reflexive and perceptual evaluation}

This study adopts a hierarchical approach to assess low-order cognitive processes at two levels:
\begin{itemize}

    \item Reflexive Evaluation: Automatic, unconscious responses are assessed using steady-state paradigms, namely SSVEP (visual) and SSAEP (auditory), which reveal basic sensory reactivity without requiring awareness. SSVEPs reflect visual processing \cite{b10} and are typically elicited by stimuli blinking at fixed frequencies. SSAEPs, evoked by amplitude modulated tones, indicate auditory system integrity and are sensitive to different states of consciousness \cite{b11}. Reflexive responses are identified by peaks at the stimulation frequency.
    
    \item Perceptual Evaluation: Perceptual processing is measured via the MMN, a pre-attentive neural response to auditory changes relying on sensory memory, which do not require selective attention. MMN has been used to assess consciousness in DoC patients and is typically elicited using oddball paradigms, where infrequent “deviant” sounds are embedded in a sequence of frequent “standard” tones. The resulting difference wave exhibits a negative peak approximately 200 ms after stimulus onset \cite{b12}.

\end{itemize}

The experimental session included four paradigms: SSVEP, SSAEP, and two variations of MMN. The respective durations were approximately 50 seconds, 1.5 minutes, 6 minutes, and 6 minutes. The following sections detail the protocol for each paradigm. The overall experimental pipeline is shown in 
Fig.\ref{fig:pipeline}.

\subsubsection{Mismatch Negativity}

\begin{figure}[htbp]
\hspace{-0.2cm}
\centerline{\includegraphics[width=0.47\textwidth]{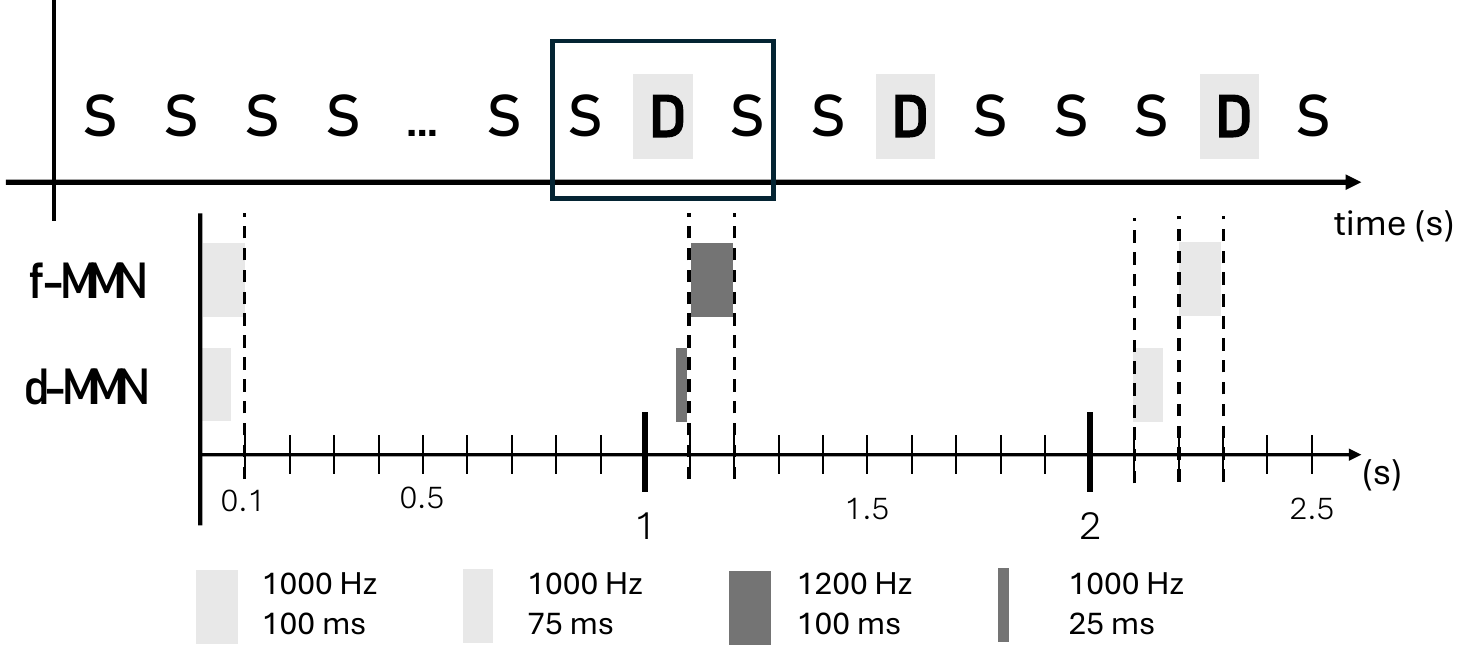}}
\caption{Mismatch Negativity Paradigms: frequency (f-MMN) and duration (d-MMN) approaches.}
\label{fig:mmn_paradigma}
\vspace{-0.5cm}
\end{figure}

Two oddball paradigms were used to elicit MMN, one based on stimulus duration (d-MMN) and the other on stimulus frequency (f-MMN). Each paradigm consisted of 300 auditory events presented in an 80:20 ratio (240 standard and 60 deviant tones). In both paradigms, the standard stimulus was a 1000 Hz pure tone, and the inter-stimulus interval (ISI) was 1000 ms (see Fig.\ref{fig:mmn_paradigma}). All sounds were generated using PsychoPy’s sound library. The sequence of tones was randomized with two constraints: (1) at least 15 standard tones were required before the first deviant tone, and (2) deviant tones were not permitted to occur consecutively.


In the f-MMN paradigm, all tones had a fixed duration of 100 ms, and deviant tones were presented at 1200 Hz. In the d-MMN paradigm, all tones had a fixed frequency of 1000 Hz, with standard tones lasting 75 ms and deviant tones shortened to 25 ms. 

\subsubsection{SSVEP}

the SSVEP paradigm employed a color alternation method, where a single rectangle alternated between black and white, against a gray background instead of the typical checkerboard stimuli. The visual flicker stimulation was defined by (\ref{eq:frames}):
\vspace{-0.2cm}
\begin{equation}
s(f_s,i) = \text{square}\left(2\pi  f_s \cdot \left(\frac{i}{\text{RefreshRate}}\right)\right)
\label{eq:frames}
\end{equation}
where \(square(\cdot)\) generates a 50\% duty cycle square wave with levels 0 and 1, and \(i\) is the frame index. The stimulus frequency \(f_s\) was set to 15 and 22 Hz. For each frequency, there were three trials of 10 seconds, preceded by a 5-second baseline period. During the baseline phase, participants had to fixate on a red dot at the center of the screen.

\subsubsection{SSAEP} 

The SSAEP paradigm consisted of two auditory stimuli synthesized using custom Matlab scripts. Both signals were modulated using Sine Amplitude Modulation, with carrier frequencies \(f_c\) of 1000 Hz and 500 Hz and corresponding modulating frequencies \(f_m\) of 41 Hz and 39 Hz \cite{b13}. The modulation depth \(m\) was set to 100\% for both tones. The resulting auditory signal, \(s(t)\), is defined as follows: 
\begin{equation}
s(t) =  m \cdot \sin(2\pi f_c t) \cdot \sin(2\pi f_m t)
\label{eq:SAM}
\end{equation}

Each stimulus was presented for 30 seconds, preceded by a 5-second silent baseline period.

\section{Results and Discussion}

\begin{figure}[t]
\centerline{\includegraphics[width=0.5\textwidth]{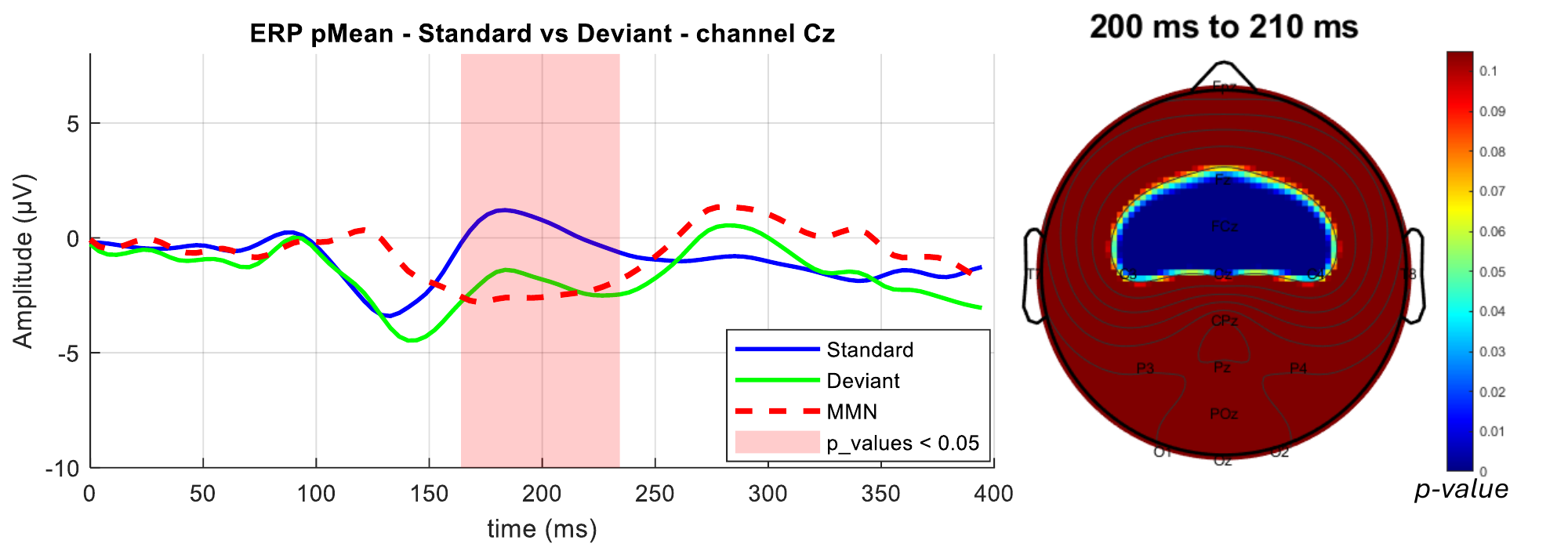}}
\vspace{-0.1cm}
\caption{(left) Grand-average waveform of the f-MMN across participants at channel Cz, with statistically significant time points \((p < 0.05)\); (right) Statistical topography from CBPT for f-MMN group analysis.}
\label{fig:f-mmn}
\vspace{-0.3cm}
\end{figure}

MMN analysis was performed in the time domain by extracting epochs from -100 ms to +400 ms, time-locked to stimulus onset. Baseline correction was performed using the -100 to 0 ms pre-stimulus window. Averages were computed separately for standard (240 epochs) and deviant (60 epochs) events. No additional preprocessing steps were applied. The one-sample Kolmogorov–Smirnov test indicated significant deviations from normality in the data distributions. Therefore, to assess significant differences between conditions, a non-parametric cluster-based permutation test (CBPT, \(p < 0.05\)) was applied. The grand average for f-MMN is shown in Fig.\ref{fig:f-mmn}, as well as the corresponding statistical topography. Support Vector Machine (SVM) classification was applied separately to each channel using 5-epoch averages within the 150–350 ms time window. A radial basis function (RBF) kernel was used, and class imbalance was addressed by applying class weight balancing. Hyperparameter tuning was performed via grid search with 5-fold cross-validation, and performance was evaluated using balanced accuracy. Results are reported in Table \ref{tab:mmn_combined} for channels Fz, FCz, Cz, CPz, and Pz, alongside amplitude and latency measures for the MMN paradigms.

\begin{figure}[t]
\centerline{\includegraphics[width=0.5\textwidth]{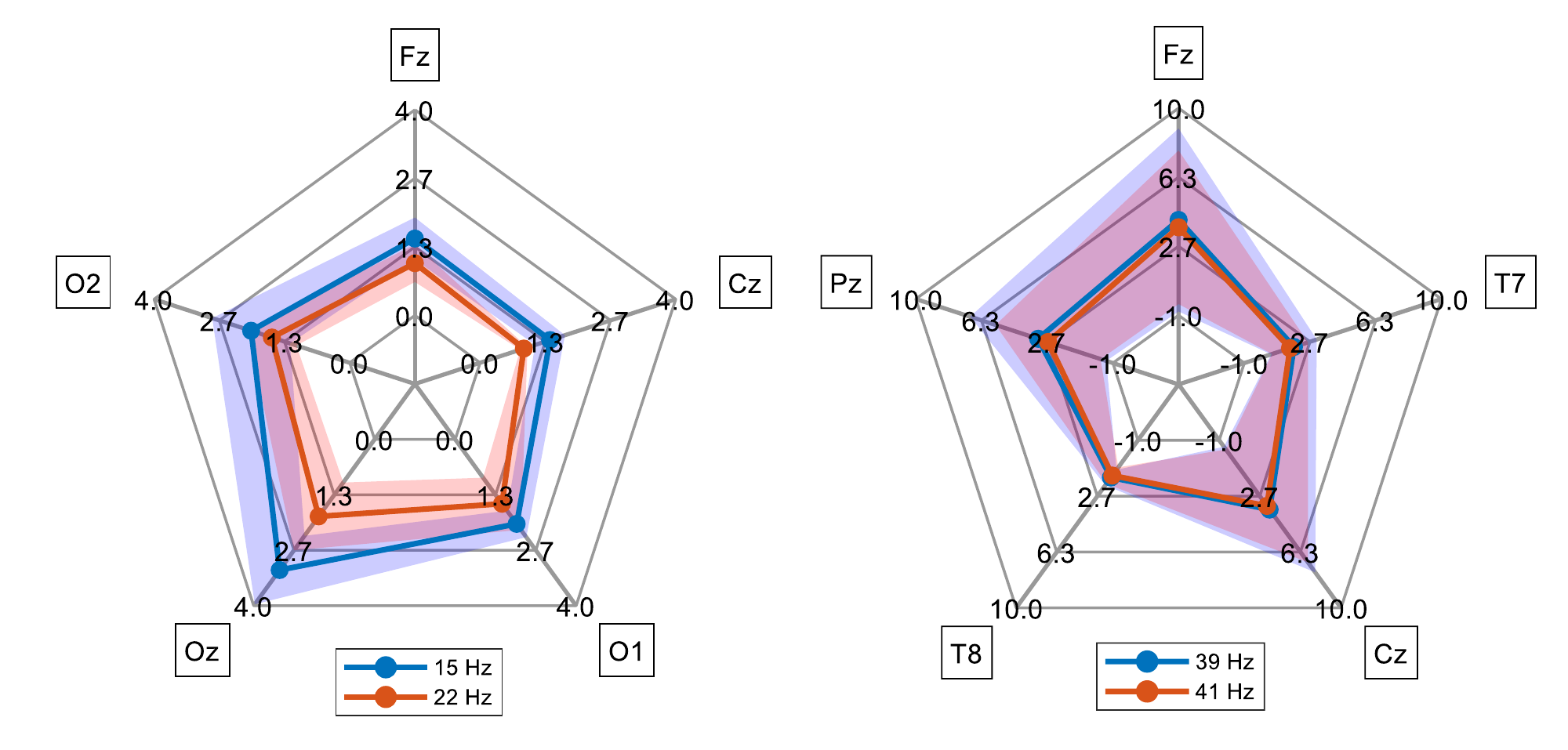}}
\caption{(left) Relative band power of SSVEP responses at 15 Hz and 22 Hz; (right) Relative band power of SSAEP responses at 39 Hz and 41 Hz. Shaded areas indicate standard deviation.}
\label{fig:ss}
\vspace{-0.5cm}
\end{figure}

Both SSVEP and SSAEP were analyzed in the frequency domain using Power Spectral Density (PSD) estimation, \(P_x(f)\), via Welch's method. The PSD was computed using an FFT with $N_{\text{FFT}} = 256$ (1 second long), yielding a spectral resolution of 1 Hz, and performed on a per-trial basis, with each epoch (10 seconds for SSVEP, 30 seconds for SSAEP, and a pre-stimulus from -1000 ms) time-locked to stimulus onset. A Hann window was applied with a 50\% overlap between consecutive segments for spectral estimation. To normalize power across participants, we used the Relative Band-Power (RBP) for SSVEP, and for SSAEP, according to (\ref{eq:RBP}):
\begin{equation}
\small
\text{RBP}(f_s) = \frac{P(f_s)}{\frac{1}{2} \left( P(f_s - 2~\text{Hz}) + P(f_s + 2~\text{Hz}) \right)}
\label{eq:RBP}
\end{equation}
where \(P(f_s)\) is the power at the stimulus frequency \(f_s\), and the denominator represents the average power of neighboring frequencies (\(f_s \pm 2\)~Hz). The grand-average results for one trial at different EEG channels are shown in Fig.\ref{fig:ss}. 

\begin{table*}[t]
\centering
\small
\caption{Comparison of MMN features across electrodes: mean amplitude (in $\mu$V), latency (in ms), and SVM balanced classification accuracy (in \%) for both f-MMN and d-MMN, averaged across five participants (mean $\pm$ standard deviation).}
\label{tab:mmn_combined}
\begin{tabular}{l|ccc|ccc}
\toprule
\textbf{Electrode} & \multicolumn{3}{c|}{\textbf{f-MMN}} & \multicolumn{3}{c}{\textbf{d-MMN}} \\
\cmidrule(lr){2-4} \cmidrule(lr){5-7}
& Amplitude ($\mu$V) & Latency (ms) & Accuracy (\%) & Amplitude ($\mu$V) & Latency (ms) & Accuracy (\%) \\
\midrule
Fz   & -5.96 $\pm$ 2.34 & 249.2 $\pm$ 102.5 & 59.4 $\pm$ 7.1  & -6.85 $\pm$ 2.59 & 199.2 $\pm$ 38.3  & 67.7 $\pm$ 9.4  \\
FCz  & -5.41 $\pm$ 2.11 & 250.8 $\pm$ 101.4 & 59.9 $\pm$ 7.1  & -6.25 $\pm$ 2.43 & 182.8 $\pm$ 65.1  & 63.4 $\pm$ 9.2  \\
Cz   & -4.85 $\pm$ 1.83 & 226.6 $\pm$ 129.6 & 61.1 $\pm$ 9.3  & -5.12 $\pm$ 2.12 & 185.2 $\pm$ 67.1  & 64.4 $\pm$ 13.9 \\
CPz  & -4.19 $\pm$ 1.37 & 252.3 $\pm$ 101.4 & 58.7 $\pm$ 12.9 & -4.26 $\pm$ 2.53 & 153.1 $\pm$ 99.0  & 66.3 $\pm$ 12.6 \\
Pz   & -3.40 $\pm$ 1.16 & 259.4 $\pm$ 94.5  & 54.7 $\pm$ 13.1 & -3.56 $\pm$ 2.45 & 154.7 $\pm$ 100.6 & 63.6 $\pm$ 10.4 \\
\bottomrule
\end{tabular}
\vspace{-0.4cm}
\end{table*}

The results demonstrate that all four paradigms successfully elicited the expected responses. SSVEP stimulation at 15 Hz elicited more prominent spectral peaks compared to 22 Hz in the occipital area (Oz), aligning with findings reported in the literature. In contrast, SSAEP responses showed comparable peak amplitudes at both tested frequencies. Despite the relatively short stimulation period, both steady-state visual and auditory paradigms produced clear responses at their respective stimulation frequencies, with no visual or auditory discomfort reported by the participants.

Regarding the MMN paradigms, group-level analysis confirmed statistically significant MMN activity for the f-MMN condition at the expected electrodes and latencies. However, at the individual level, f-MMN peaks were less consistent. For the d-MMN condition, MMN-like responses were observed in most participants, with statistically significant effects and classification accuracy reaching up to 80\%. These findings confirm the presence of detectable MMN responses in both paradigms. Although inter-individual variability poses a challenge, it may be partly attributed to the small samples size. Nonetheless, the results suggest that both paradigms are effective and capable of reliably eliciting MMN responses.

The ultimate goal of this work is to build a neurophysiological report to assess whether new patients fit within the established norms. To determine this, a z-score-based method is applied according to (\ref{eq:z}):

\vspace{-2mm}
\begin{equation}
\small
z = \frac{X_{\text{patient}} - \mu_{\text{norm}}}{\sigma_{\text{norm}}}
\label{eq:z}
\end{equation}
\vspace{-2mm}

\begin{figure}[htbp]
\centerline{\includegraphics[width=0.45\textwidth]{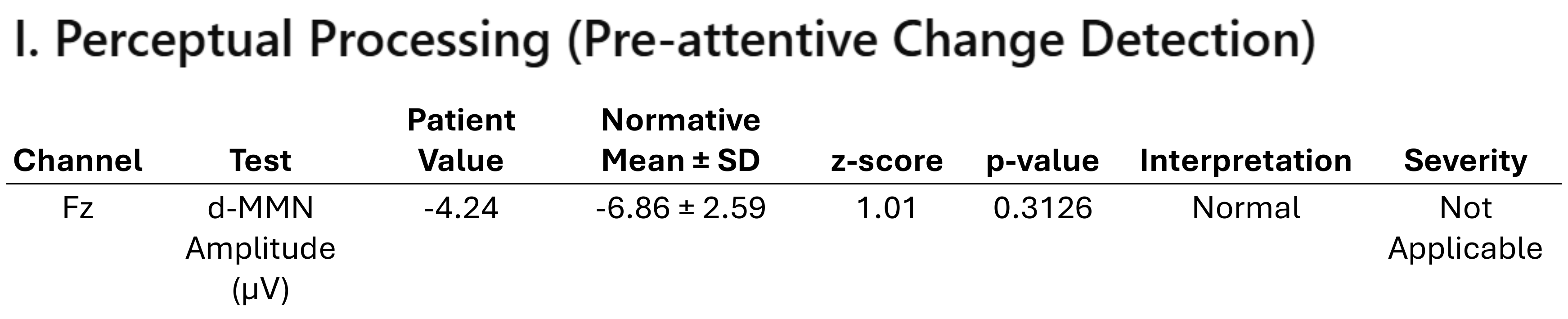}}
\caption{Example of a neurophysiological report entry for assessing perceptual processing using the d-MMN paradigm, based on amplitude analysis at channel Fz in a healthy volunteer.}
\label{fig:report}
\end{figure}

Reference ranges are defined as the mean ±1.96 standard deviations of the normative control group, corresponding to a 95\% confidence interval under the assumption of normal distribution. Values within the statistical normal range (mean ±1.96 SD) are considered normal, while those outside are considered abnormal. Additionally, p-values derived from the z-scores provide a certainty of the severity, which can be provided using \( p = 2 \cdot (1 - \Phi(|z|)) \), where \( \Phi(|z|) \) is the cumulative distribution function of the standard normal distribution. Fig.\ref{fig:report} presents an example application of the proposed evaluation method.
\vspace{-0.3cm}

\vspace{-0.5cm}
\section{Conclusion}

The proposed reflexive and perceptual evaluation framework successfully elicited reflexive and perceptual neural responses across all paradigms, demonstrating its potential for assessing cognitive function in CLIS patients. While the results are promising, the observed variability in MMN responses, likely due to the small sample size, poses challenges for establishing robust normative markers. Incorporating additional metrics (e.g., PLE, power bands) may improve assessment accuracy; however, expanding the control group is essential. This expansion, already underway, will help refine normative benchmarks and enhance the reliability and generalizability of the neurophysiological report. Future work includes clinical validation with CLIS and DoC populations.

\bibliographystyle{IEEEtran}
\bibliography{refs.bib}

\end{document}